\documentclass[sigconf]{acmart}

\AtBeginDocument{%
  }

\setcopyright{acmlicensed}
\copyrightyear{2018}
\acmYear{2018}
\acmDOI{XXXXXXX.XXXXXXX}

\acmConference[Conference acronym 'XX]{Make sure to enter the correct
  conference title from your rights confirmation emai}{June 03--05,
  2018}{Woodstock, NY}
\acmISBN{978-1-4503-XXXX-X/18/06}

\usepackage{enumitem}
\usepackage{multirow}
\usepackage{diagbox}

\usepackage{natbib}
\setcitestyle{numbers,square}
\usepackage{bm}

\usepackage{hyperref}
\usepackage{microtype}
\usepackage{graphicx}
\usepackage{subfigure}
\usepackage{booktabs}
\usepackage{amsmath}
\usepackage{algorithm,algorithmic}
\begin{document}

\title{RLMiner: Finding the Most Frequent k-sized Subgraph via Reinforcement Learning}

\author{Wei Huang}
\affiliation{%
  \institution{University of New South Wales}
  \country{Australia}
}
\email{w.c.huang@unsw.edu.au}

\author{Hanchen Wang}
\affiliation{%
  \institution{University of Technology Sydney}
  \country{Australia}
}
\email{Hanchen.Wang@uts.edu.au}

\author{Dong Wen}
\affiliation{%
  \institution{University of New South Wales}
  \country{Australia}
}
\email{dong.wen@unsw.edu.au}

\author{Xin Cao}
\affiliation{%
  \institution{University of New South Wales}
  \country{Australia}
}
\email{xin.cao@unsw.edu.au}

\author{Bocheng Han}
\affiliation{%
  \institution{Vecton AI}
  \country{Australia}
}
\email{bocheng.han@vectonai.com}

\author{Ying Zhang}
\affiliation{%
  \institution{Zhejiang Gongshang University}
  \country{China}
}
\email{ying.zhang@zjgsu.edu.cn}

\author{Wenjie Zhang}
\affiliation{%
  \institution{University of New South Wales}
  \country{Australia}
}
\email{wenjie.zhang@unsw.edu.au}

\begin{abstract}
Identifying the most frequent induced subgraph of size $k$ in a target graph is a fundamental graph mining problem with direct implications for Web-related data mining and social network analysis. 
Despite its importance, finding the most frequent induced subgraph remains computationally expensive due to the NP-hard nature of the subgraph counting task.
Traditional exact enumeration algorithms often suffer from high time complexity, especially for a large graph size $k$. 
To mitigate this, existing approaches often utilize frequency measurement with the Downward Closure Property to reduce the search space, imposing additional constraints on the task. 
In this paper, we first formulate this task as a Markov Decision Process and approach it using a multi-task reinforcement learning framework.
Specifically, we introduce RLMiner, a novel framework that integrates reinforcement learning with our proposed task-state-aware Graph Neural Network to find the most frequent induced subgraph of size $k$ with a time complexity linear to $k$.
Extensive experiments on real-world datasets demonstrate that our proposed RLMiner effectively identifies subgraphs with frequencies closely matching the ground-truth most frequent induced subgraphs, while achieving significantly shorter and more stable running times compared to traditional methods. The source code is available at \url{https://anonymous.4open.science/r/RLMiner-CD2C/}.
\end{abstract}



\maketitle

\section{Introduction}
\begin{figure}[t]
     \centering
     \resizebox{0.8\linewidth}{!}{\includegraphics[scale=1]{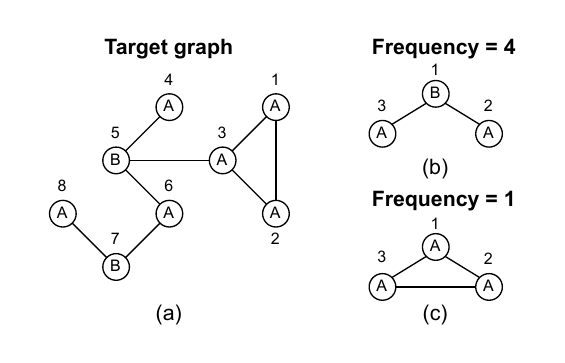}}
     \caption{An example of frequent subgraph mining. (a) A target graph. (b) and (c) Induced subgraphs of the target graph.}
     \vspace{-5mm}
     \label{fig:example}
 \end{figure}

Finding the most frequent induced subgraph in a labeled undirected target graph is a crucial but challenging step in characterizing the complex structure property of graphs, it has significant implications across various domains, from bioinformatics \cite{cho2009predicting,agrawal2018large} and chemistry \cite{cereto2015molecular} to social network analysis \cite{janssen2012model,leskovec2010signed}. The task is defined as finding the most frequent induced subgraph of size $k$ given a undirected labeled target graph. As shown in Figure \ref{fig:example}, subgraph (b) is the most frequent induced subgraph of size $3$ in target graph (a) with frequency of $4$. 

 Given its wide applications, there have been continuing interests in tackling this problem. Traditional exact enumeration approaches often enumerate and count the frequency of all possible connected subgraphs of size $k$ \cite{paredes2013towards,hovcevar2014combinatorial,wernicke2006efficient,kashani2009kavosh}. However, counting the frequency of a subgraph was proven to be NP-hard \cite{ribeiro2021survey}, and the number of possible subgraphs (search space) grows exponentially as $k$ increases. Thus, traditional approaches often suffer from high computational complexity even for small and sparse graphs. One popular way to prune the search space is to use frequency measurements that satisfy Downward Closure Property (DCP) \cite{fiedler2007subgraph,bringmann2008frequent,kuramochi2005finding}, such as \textit{minimum image based support (MNI)} \cite{bringmann2008frequent}, which imposes additional restrictions and limits the applicability of the task . 
 There are also methods \cite{yan2002gspan,gaston} for discovering the frequent subgraphs on labeled graphs without DCP restriction, but those works focused on the subgraph frequency across graph dataset instead of a single target graph, where the frequency of a subgraph is either counted as 0 or 1 in a single target graph. 
 Note that, a graph dataset can be considered as multiple connected components of a single graph without overlapping, thus it is often easier to identify frequent subgraphs in a graph dataset than a single graph~\cite{elseidy2014grami}.

To address the limitations of traditional methods, recent years have seen efforts to leverage Graph Neural Networks (GNNs) to tackle this problem. 
Ying et al.~\cite{ying2024representation} has proposed the first deep learning-based approach SPMiner to find frequent induced subgraphs of size $k$ in a target graph based on order embedding space. 
However, it focused on unlabeled graphs and also employed frequency measurements with Downward Closure Property. 
Thus, there remains a lack of research focused on identifying frequent subgraphs within a single labeled graph without imposing the DCP restriction on frequency measurements.

Motivated by the limitations of existing approaches, we propose a novel efficient way to find the most frequent induced subgraph of size $k$ in a single labeled target graph without DCP restriction on frequency measurement. 
First, we model this problem as a multi-task reinforcement learning problem, where each subgraph size $k$ represents a single task.
Each task is formulated as a Markov Decision Process (MDP). 
Next, we present RLMiner, a reinforcement learning algorithm that combines Soft Actor Critic (SAC) with our proposed task-state-aware Graph Neural Network to efficiently search for the most frequent induced subgraph. 
Our proposed approach avoids time-consuming subgraph enumeration and subgraph counting during inference, offering an approximate solution with a time complexity linear to the subgraph size $k$. 

Extensive experiments on real-world datasets demonstrate that our proposed $RLMiner$ can solve the problem with high performance while the running time of exact enumeration is much longer (more than $2000\times$ on harder target graphs) than the running time of RLMiner. Our contribution can be summarized as follows:

\begin{itemize}[leftmargin=*]
    \item To the best of our knowledge, RLMiner is the first approximation method that finds the most frequent induced subgraph of size $k$ in a single labeled target graph without DCP restriction on frequency measurement.
    \item Our proposed RLMiner leverages reinforcement learning, and can efficiently approximate the most frequent subgraph with a time complexity linear to $k$.
    \item We propose a task-state-aware GNN architecture for multi-task reinforcement learning on graphs that incorporates the task and state information into message-passing.
    \item Extensive experiments demonstrate that our proposed RLMiner can achieve high approximation ratio that is close to the solution found by exact enumeration with a much shorter running time.
\end{itemize}

\section{Related Work}
\label{sec:related}
\noindent \textbf{Traditional Approaches for Frequent Subgraph Mining.}
ESU \cite{wernicke2006efficient} and Kavosh \cite{kashani2009kavosh} are two of the well-known exact enumeration methods for frequent subgraph mining on unlabeled graphs, they both
work by enumerating all occurrences of $k$-sized subgraph, then apply an isomorphic test to each occurrence. Later on, FASE \cite{paredes2013towards} was proposed based on ESU, which integrates the isomorphism tests into the enumeration. However, these methods suffered from high time complexity with exponential search space, and they were designed only for unlabeled target graphs. To effectively reduce the search space of subgraph enumeration, several methods \cite{elseidy2014grami,yuan2023t,jiang2021efficient} proposed to use frequency measurement with Downward Closure Property to find frequent subgraphs on a labeled target graph. Furthermore, there are some other works focused on frequent subgraph mining on graph database \cite{yan2002gspan,gaston}, which only considered whether a subgraph appears in a single graph or not, instead of the subgraph frequency in a single graph.

\noindent \textbf{Traditional Approaches for Subgraph Counting.}
The frequent subgraph mining problem heavily relies on subgraph counting. Existing traditional approaches for subgraph counting mainly based on backtracking, where partial solutions are recursively extended. Ullmann \cite{ullman1976algorithm} proposed the first algorithm for subgraph counting based on backtracking. Later on, several in-memory subgraph counting methods have been proposed with backtracking technique as summarized in \cite{sun2020memory}, such as VF3 \cite{carletti2017challenging}, QuickSI \cite{shang2008taming} and GraphQL \cite{he2008graphs}. 

\noindent \textbf{Deep Learning Approaches for Frequent Subgraph Mining.} 
Recently, SPMiner \cite{ying2024representation} proposed the first deep learning approach for frequent subgraph mining, aiming to search for the most frequent $k$-sized induced subgraphs using order space embeddings. However, it focused on unlabeled target graphs with frequency measurement satisfies Downward Closure Property. For our problem, we aim to find the most frequent $k$-sized induced subgraph on labeled graphs based on the number of unique isomorphisms, which is more intuitive but also more challenging than frequency measurement with Downward Closure Property. To the best of our knowledge, this problem has not been explored in prior work.

\section{Preliminaries}
\label{sec:preliminaries}
\subsection{Problem Setup}
In this work, we focus on finding the most frequent induced subgraphs on undirected labeled target graphs, the problem is defined as follows:

\noindent \textbf{Problem Statement.} \textit{Given a subgraph size $k$, i.e., number of nodes and an undirected labeled target graph $G^T = (V^T,E^T,L^T)$ consists of a set of nodes $V^T$, a set of edges $E^T$ and a labeling function $L^T$ that assigns each node a label, find an induced subgraph of size $k$ from $G^T$ with the highest frequency.}

\noindent In our problem, the frequency of a subgraph $G^S$ in $G^T$ is defined as follows:

\noindent \textbf{Definition 1. \textit{Subgraph Frequency.}} \textit{The frequency of a subgraph $G^S$ in $G^T$ is the number of unique induced subgraph from $G^T$ that exists an isomorphism to $G^S$.}

\noindent \textbf{Definition 2. \textit{Induced Subgraph.}} \textit{$G^S = (V^S,E^S,L^S)$ is a subgraph of $G^T = (V^T,E^T,L^T)$ if $V^S \subseteq V^T$ and $E^S \subseteq E^T$. $G^S$ is an induced subgraph if $E^S$ consists of all edges in $E^T$ with endpoints in $V^S$.}

\noindent \textbf{Definition 3. \textit{Graph Isomorphism.}} \textit{Two graphs $G = (V,E,L)$ and $G' = (V',E',L')$ are isomorphic if and only if there exists a bijection $f: V' \xrightarrow{} V$ such that $L'(v') = L(f(v'))$ for all nodes $v' \in V'$, and $(u',v')\in E'$ if and only if $(f(u'),f(v'))\in E$.}

 Note that, two induced subgraphs of $G^T$ isomorphic to $G^S$ are considered different only if they have at least one node they do not share. 
For instance, in Figure \ref{fig:example}, there are $6$ isomorphism mappings of subgraph (c) in target graph (a): 

\noindent$\{(v^T_1,v^T_2,v^T_3),(v^T_1,v^T_3,v^T_2),(v^T_2,v^T_1,v^T_3),(v^T_2,v^T_3,v^T_1),(v^T_3,v^T_1,v^T_2),\\(v^T_3,v^T_2,v^T_1)\},$

\noindent where $v^T_i$ denotes the node in target graph (a). However, these mappings all share the same node sets $\{v^T_1,v^T_2,v^T_3\}$, thus we count the frequency of subgraph (c) in target graph (a) as $1$. Obviously, this is more intuitive than using a frequency count of $6$. Following the same principle, subgraph (b) has a frequency count of $4$ in target graph (a).

 \begin{figure*}[ht]
     \centering
     \resizebox{0.85\textwidth}{!}{\includegraphics[scale=1]{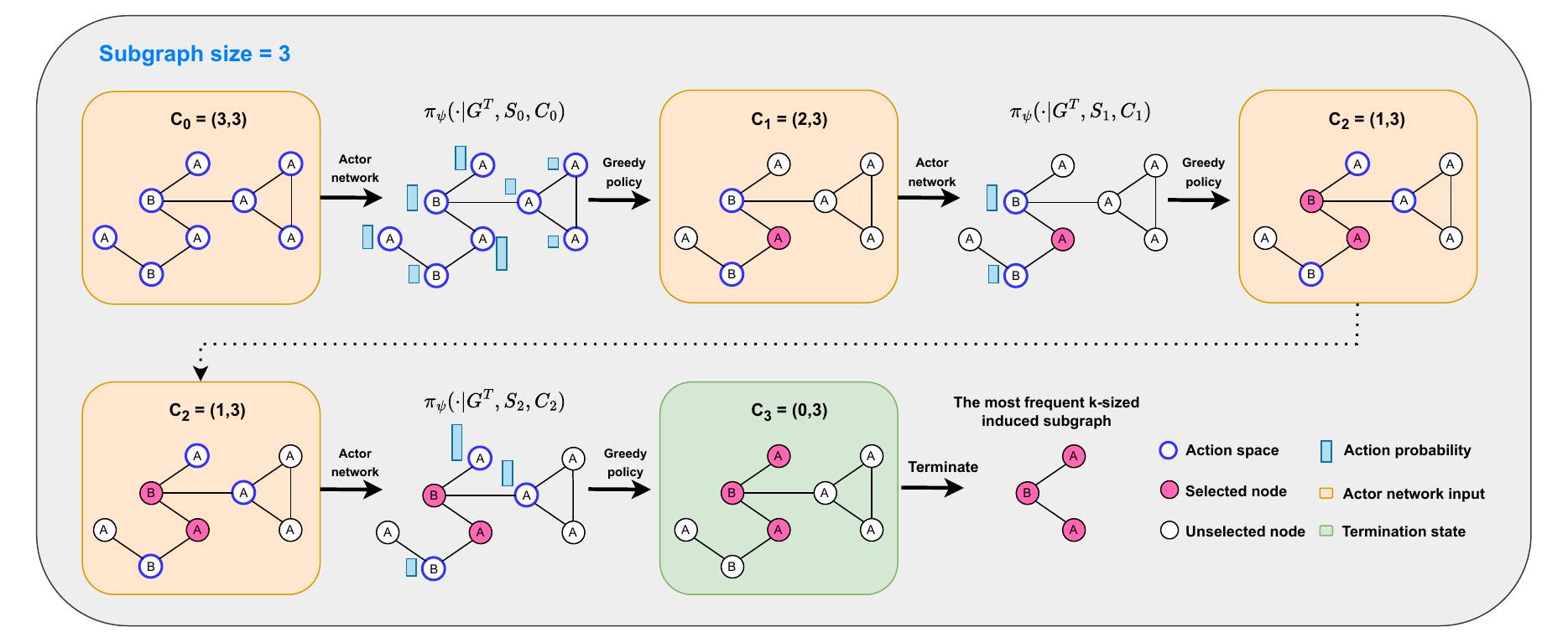}}
     \vspace{-3mm}
     \caption{An overview of RLMiner inference process. In each time step, $RLMiner$ computes the action probability distribution using the Actor network, then select the action with highest probability according to greedy policy.}
     \vspace{-2mm}
     \label{fig:rlminer}
 \end{figure*}
\subsection{Downward Closure Property Frequency Measure}
To illustrate the inconsistency of using frequency measure with DCP restriction, we present the following example using \textit{MNI} measurement since it is the most popular frequency measurement used in existing works:

\noindent \textbf{Definition 4. \textit{MNI.}} \textit{Let $f_1,...,f_m$ be the set of isomorphisms of a subgraph $G^S = (V^S,E^S,L^S)$ in a target graph $G^T = (V^T,E^T,L^T)$. Let $F(v^S)=\{f_1(v^S),...,f_m(v^S)\}$ be the set that contains the distance nodes in $G^T$ whose functions $f_1,...,f_m$ map a node $v^S \in V^S$. The minimum image based support (MNI) of $G^S$ in $G^T$ is defined as $min\{|F(v^S)| \mid \forall v^S\in V^S\}$.}

Consider the target graph (a) and subgraph (b) in Figure \ref{fig:example}, we have:

\noindent$F(v^S_1) = \{v^T_5,v^T_7\}$, $F(v^S_2)=\{v^T_3,v^T_4,v^T_6,v^T_8\}$, $F(v^S_3)=\{v^T_3,v^T_4,v^T_6,v^T_8\}$

\noindent where $v^S_i$ are the nodes in subgraph (b) and $v^T_i$ are the nodes in target graph (a), the \textit{MNI} of subgraph (b) in target graph (a) is $min\{2,4,4\}=2$. However there are 4 distinct induced subgraphs in (a) that are isomorphic to subgraph (b):

\noindent$\{(v^T_3,v^T_4,v^T_5),(v^T_3,v^T_5,v^T_6),(v^T_4,v^T_5,v^T_6),(v^T_6,v^T_7,v^T_8)\}$

Obviously, a frequency count of $4$ is more intuitive than a frequency count of $2$. Also, the \textit{MNI} of subgraph (c) is $3$, which has a higher \textit{MNI} than subgraph (b). Therefore, in our problem, we use the most straightforward frequency measurement without DCP restriction as defined in Definition 1. 

Note that, in the above example, there are at most $2$ nodes with label ``B" in target graph (a), thus the \textit{MNI} of subgraph (b) is always bounded by $2$, such Downward Closure Property (DCP) allows traditional methods to prune a significant amount of search space. However, our proposed RLMiner does not rely on the frequency measurement with DCP to prune the search space, it can approximately find the most frequent subgraph with time complexity linear to $k$.

\section{Proposed Approach}
\label{sec:approach}
In this section, we present our RLMiner that utilizes multi-task reinforcement learning to find the most frequent induced subgraph of size $k$. The overall search process of RLMiner is illustrated in Figure \ref{fig:rlminer}. 

\subsection{Markov Decision Process Formulation}
\label{sec:mdp}
Given a set of subgraph sizes $K$, finding the most frequent induced subgraph of size $k \in K$ can be viewed as a multi-task learning problem where each size $k$ represents a single task, and can be formulated as the following MDP: 

    \noindent\textbf{State \& Task:} Given a target graph $G^T = (V^T,E^T,L^T)$, the state $S_t$ at time $t$ can be defined as $\bm{x}_t$, where $\bm{x}_t$ is a vector such that $\bm{x}_{t,i} = 1$ if $v^T_i \in V^T$ has been selected in the past, otherwise $\bm{x}_{t,i} = 0$. Task at time $t$ can be represented as $C_t = (k-t,k)$, where the first element indicates the number of remaining nodes required to be selected into the current solution, and the second element indicates the size of the subgraph we aim to find. Incorporating the number of remaining steps offers additional task-related information and helps to distinguish successive states. 
    
    \noindent\textbf{Action:} The action $a_t$ at time $t$ is defined as selecting an unselected node $v^T_{a_t} \in V^T$ such that $\bm{x}_{t,{a_t}} = 0$, and $v^T_{a_t}$ is connected to at least one of the selected nodes. All the nodes can be selected for the initial state $S_0$, since none of the nodes have been selected. The valid action space at time $t$ is denoted as $\mathcal{A}_t$:
    
    \begin{equation}
    \small
        \mathcal{A}_t = 
        \begin{cases}
            \{{a_t} | v^T_{a_t} \in V^T\}  &\text{if } t = 0 \\
            \{{a_t} | \{v^T_{a_t},v^T_{j}\} \in V^T, \bm{x}_{t,{a_t}} = 0, \bm{x}_{t,j} = 1, (v^T_{a_t},v^T_{j}) \in E^T \} &\text{else}
        \end{cases}
    \end{equation}
    
    \noindent\textbf{State \& Task transition:} After selecting a node $v^T_{a_t}$, we will set $\bm{x}_{t+1,{a_t}} = 1$. Also, we will update the task to $C_{t+1} = (k-t-1,k)$.
    
    \noindent\textbf{Termination:} Given a subgrpah size $k$, a termination state is defined as a state with the number of selected nodes equals to $k$, such that $C_t = (0,k)$.
    
    \noindent\textbf{Reward:} Given a subgraph size $k$, the reward $r((G^T,S_t,C_t),a_t)$ of taking an action $a_t$ given target graph $G^T$, state $S_t$ and task $C_t$ is always 0 for all $t < k-1$, that is, the reward is always 0 until reaching the termination state. 
    
    For $a_{k-1}$ that results in termination state, we now have our $k$-sized subgraph $G^S_k$ induced  by the selected nodes $\{v^T_i \mid v^T_i \in V^T, \bm{x}_{k,i} = 1 \}$ at termination, the most straightforward and common way is to set the reward 
    \begin{equation}
    \small
    \label{eq:reward1}
        r((G^T,S_{k-1},C_{k-1}),a_{k-1}) = \textit{frequency }(G^S_k,G^T)
    \end{equation}
    where \textit{frequency }$(G^S_k,G^T)$ is the frequency of induced subgraph $G^S_k$ in $G^T$ as defined in Definition 1. 
    
    However, the scale of the subgraph frequency varies across target graphs according to factors such as target graph size, target graph edge density, as well as subgraph size $k$. 
    This poses challenges for accurately estimating Q-values, particularly when RLMiner is generalized to unseen target graphs. 
    Therefore, the reward is set as follows:
    \begin{equation}
    \small
    \label{eq:reward2}
        r((G^T,S_{k-1},C_{k-1}),a_{k-1}) = \frac{\textit{frequency }(G^S_k,G^T)}{\textit{frequency }({G^S_k}^*,G^T)}
    \end{equation}
    where ${G^S_k}^*$ is the most frequent induced subgraph found by the exact enumeration. In this way, $r(S_{k-1},a_{k-1})$ always lies within $(0,1]$ regardless of the graph size, graph density or subgraph size $k$. 
    Note that, the reward is only needed in the training stage, thus
    ${G^S_k}^*$ only needs to be obtained for the training graphs within the pre-processing step. 
  
    However, for large graphs, dense graphs, and even small sparse graphs with large $k$, obtaining ${G^S_k}^*$ is computationally expensive and often infeasible.
    Hence, we introduce an alternative way to normalize the reward as follows:
    \begin{equation}
    \small
    \label{eq:reward3}
    \begin{aligned}
        r((G^T,S_{k-1},C_{k-1}),a_{k-1}) &= \frac{\textit{frequency }(G^S_k,G^T)}{|V^T|\times \text{edge density} \times k} \\
        \text{edge density} &= \frac{2|E^T|}{|V^T|\times(|V^T|-1)}
    \end{aligned}
    \end{equation}
    where we approximate $\textit{frequency }({G^S_k}^*,G^T)$ by the target graph size, target graph edge density and subgraph size $k$. 
    A detailed evaluation of reward settings of Equation \ref{eq:reward1}, Equation \ref{eq:reward2}, and Equation \ref{eq:reward3} is provided in Section \ref{sec:experiment}.
    
    \noindent\textbf{Policy}: Given an action probability distribution denoted as \\$\pi(\cdot | G^T,S_t,C_t)$. During training, RLMiner samples an action $a^*_t$ from $\pi(\cdot | G^T,S_t,C_t)$. During inference, RLMiner follows a deterministic greedy policy where it selects the best action $a^*_t$ with the highest probability such that $a^*_t = \underset{a_t \in \mathcal{A}_t}{\arg\max} \,
  \pi(a_t | G^T,S_t,C_t)$.

\begin{figure*}[ht]
     \centering
     \includegraphics[scale=0.47]{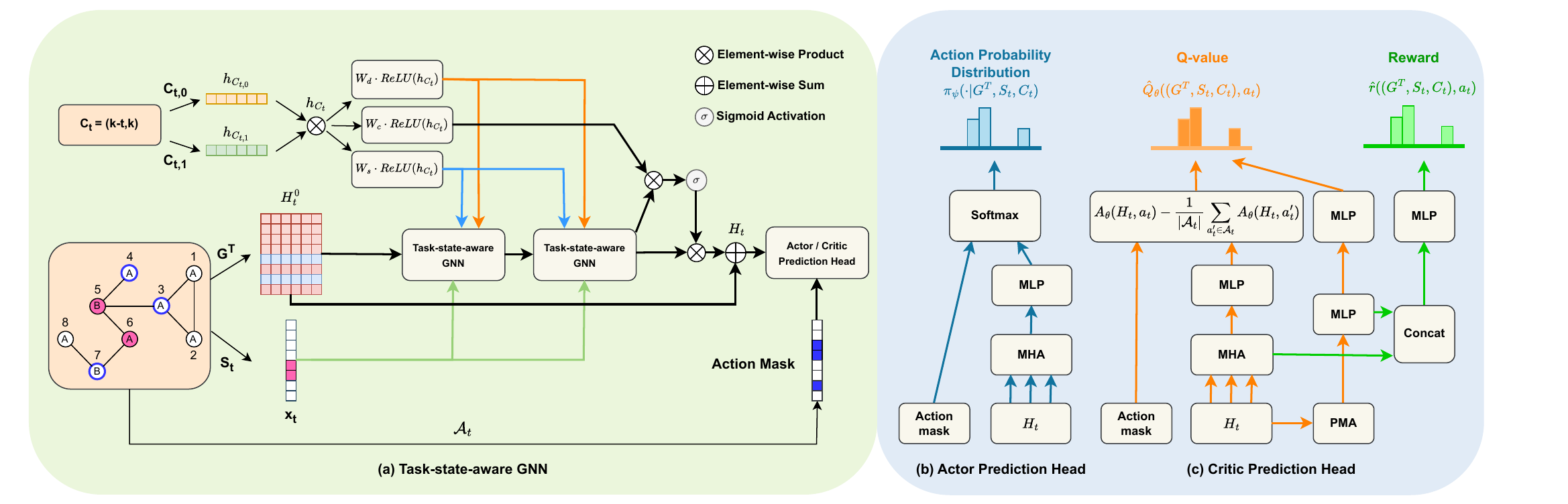}
     \vspace{-5mm}
     \caption{Architecture of the Actor network and the Critic network. (a) An overview of the proposed Task-state-aware GNN. (b) An overview of the Actor network prediction head. (c) An overview of the Critic network prediction head.}
     \vspace{-3mm}
     \label{fig:gnn}
    
\end{figure*}

\subsection{Multi-task Reinforcement Learning of RLMiner}
\label{sec:representation}
As mentioned previously, the search space of frequent subgraph mining grows exponentially according to multiple factors. 
Moreover, the NP-hard subgraph counting task is an essential step to compute the reward when training the reinforcement learning algorithm. 
To enhance both the sample efficiency and the data efficiency, we adopt the discrete action version of Soft Actor Critic (SAC)~\cite{haarnoja2018soft,haarnoja2018soft2,christodoulou2019soft} as the backbone RL algorithm for our RLMiner.

SAC consists of the Actor network and the Critic network, where the Actor network computes the probability of taking each action, and the Critic network computes the Q-value of each action. The action is selected based on the action probability distribution computed by the Actor network.

To select the best action node for a given target graph $G^T$, state $S_t$ and task $C_t$, one intuitive way is to encode the local graph structure for each node using the graph neural network, where each node's local graph structure captures several overlapping $k$-sized subgraph. 
Then, the node is selected, whose local graph structure has the highest probability of capturing the most frequent $k$-sized subgraph.
Note that, for different subgraph size $k$, the most frequent induced subgraph can have a different graph structure, i.e., the most frequent induced subgraph of size $k$ is not always a super-graph of the most frequent induced subgraph of size k' with $k' < k$. 
Thus, it is necessary to encode different local graph structures according to the value of $k$.
Furthermore, given a state $S_t$, where certain nodes have been incorporated into the partial solution, the optimal action should be chosen based on this partial solution. 
Consequently, local graph structures must be encoded differently depending on the specific partial solution.

To fulfill these requirements, both the Actor Network and Critic Network of our RLMiner consist of two components as represented in Figure~\ref{fig:gnn}: \textbf{(1)} A task-state-aware graph neural network that extracts local subgraph structures from the target graph according to the state and the task; \textbf{(2)} A prediction head that outputs the action value or probability based on the frequency of captured subgraphs.

Note that, both the Actor Network and the Critic Network use the same graph neural network architecture but with different prediction head architectures. 
Moreover, there are no learnable parameters shared between the two networks.

\subsubsection{Task-state-aware GNN} 
Before introducing our task-state-aware GNN, it is essential to introduce task-aware message-passing and state-aware message-passing.
The task-aware message-passing focuses on extracting the local graph structure for each node $v^T_i$ according to the task information $C_t$, while the state-aware message-passing focuses on extracting the local graph structure for each node $v^T_i$ according to the state information $S_t$.

\noindent \textbf{Message-passing}\quad In our proposed method, we use Local Extrema Convolution layer (LEConv) \cite{ranjan2020asap} as our basic message-passing layer since it can capture local graph structure in a simple way with low time complexity. 
LEConv performs message-passing depending on the difference between a node and its neighbors:
\begin{equation}
\small
    \bm{h}^{l+1}_{i,t} = ReLU({\bm{W}^l_1}{\bm{h}^l_{i,t}}+\sum_{j\in \mathcal{N}(i)} ({\bm{W}^l_2}{\bm{h}^l_{i,t}} - {\bm{W}^l_3}{\bm{h}^l_{j,t}}))
\end{equation}
where $\bm{h}^l_{i,t}$ denotes the representation of node $v^T_i \in V^T$ at message-passing layer $l$ and time $t$, $\mathcal{N}(i)$ denotes the neighbors of node $v^T_i$ in $G^T$, and $\bm{W}^l_1,\bm{W}^l_2,\bm{W}^l_3$ are learnable parameters at layer $l$.

\noindent \textbf{Task-aware message-passing}\quad For our problem, there exhibits a strong correlation between each task, therefore most knowledge could be shared across tasks.
To take advantage of this, we map each task to a representation and take the task representation as an additional input of the Actor network and Critic network.
The model extracts task-specific local graph structures by aggregating neighbors for each node based on the given task representation.
In this way, all network parameters are shared among all tasks, which facilitates knowledge sharing between tasks and enhances data efficiency compared to employing task-specific layers (or parameters). 
This is particularly critical for our problem, as discussed previously.

As described in Section \ref{sec:mdp}, we represent the task at time $t$ as $C_t = (k-t,k)$. We map $C_t$ to the task representation as follows:
\begin{equation}
\small
\label{eq:task_representation}
\begin{aligned}
    \bm{h}_{C_t,0} &= {\bm{W}_a} \cdot {OneHot(k-t)}, \quad  \bm{h}_{C_t,1} = {\bm{W}_b} \cdot {OneHot(k)}\\
    \bm{h}_{C_t} &= \bm{h}_{C_t,0} \odot \bm{h}_{C_t,1}
\end{aligned} 
\end{equation}
where $OneHot$ is the one-hot encoding function, $\bm{W}_a$ and $\bm{W}_b$ are learnable parameters, we map the elements of $C_t$ to representations $\bm{h}_{C_t,0}$ and $\bm{h}_{C_t,1}$, $\odot$ denotes the element-wise product, and $\bm{h}_{C_t}$ is the task representation.

The key step to encode different local graph structures of a node for different $k$ is to aggregate neighbours based on the values of $k$. 
We integrate task representation into the neighbor aggregation component of LEConv by gating mechanism:
\begin{equation}
\small
\label{eq:task_aware}
    \begin{aligned}
    {\bm{\hat{h}}^l_{i,t}} &= {\bm{W}^l_2}{\bm{h}^l_{i,t}}, \quad  
    {\bm{\hat{h}}^l_{j,t}} = {\bm{W}^l_3}{\bm{h}^l_{j,t}}\\
     {\bm{g}^l_{i,t}} &= \sigma ({\bm{W}^l_4} \cdot({\bm{h}^l_{i,t}}\odot({\bm{W}_d}ReLU({\bm{h}_{C_t}}))))\\
     {\bm{g}^l_{j,t}} &= \sigma ({\bm{W}^l_5} \cdot({\bm{h}^l_{j,t}}\odot({\bm{W}_s}ReLU({\bm{h}_{C_t}}))))\\
    \bm{h}^{l+1}_{i,t} &= ReLU({\bm{W}^l_1}{\bm{h}^l_{i,t}}+\sum_{j\in \mathcal{N}(i)}
    ({\bm{\hat{h}}^l_{i,t}} \odot {\bm{g}^l_{i,t}}  - {\bm{\hat{h}}^l_{j,t}} \odot {\bm{g}^l_{j,t}}))
   \end{aligned}
\end{equation}
where $\sigma$ denotes the element-wise sigmoid function, $\bm{W}_d$ and $\bm{W}_s$ are learnable parameters shared across all layers. Now, neighbors are aggregated based on the interaction of the source node and the destination node with the task representation.

 \noindent \textbf{State-aware message-passing}\quad As described in Section \ref{sec:mdp}, $S_t$ is represented by a vector $\bm{x}_t$. The most common way to integrate $S_t$ with the input target graph is to concatenate $\bm{x}_t$ with the node label matrix of the input graph, then simply pass into GNN. However, this approach is not feasible for our problem, when counting the frequency of subgraphs each extracted local graph structure captures, each local graph structure's representation should only contain the original node label information, letting $\bm{x}_t$ become an additional node label will lead to bias, $\bm{x_t}$ should only participate in neighbor aggregation similar to the task representation.

To aggregate neighbor according to $\bm{x}_t$, we follow a similar approach as Equation \ref{eq:task_aware}:
\begin{equation}
\small
\label{eq:state_aware}
    \begin{aligned}
        \bm{h}^{l}_{i,\bm{x}_t} &= {\bm{W}^l_4} \cdot OneHot({\bm{x}_{t,i}}), \quad 
        \bm{h}^{l}_{j,\bm{x}_t} = {\bm{W}^l_5} \cdot OneHot({\bm{x}_{t,i}})\\
        {\bm{\hat{h}}^l_{i,t}} &= {\bm{W}^l_2} \cdot ({\bm{h}^l_{i,t}} \odot \bm{h}^{l}_{i,\bm{x}_t}) \quad 
        {\bm{\hat{h}}^l_{j,t}} = {\bm{W}^l_3} \cdot ({\bm{h}^l_{j,t}} \odot \bm{h}^{l}_{i,\bm{x}_t})\\
        \bm{h}^{l+1}_{i,t} &= ReLU({\bm{W}^l_1}{\bm{h}^l_{i,t}}+\sum_{j\in \mathcal{N}(i)} (\bm{\hat{h}}^l_{i,t} - {\bm{\hat{h}}^l_{j,t}}))
    \end{aligned}
\end{equation}
Different from Equation \ref{eq:task_aware}, the above approach scales the importance of each neighbor differently according to the state of the source node and the state of the destination nodes (equivalent to the state of the edge). Therefore, the gating mechanism is not required for state-aware message-passing.

\noindent \textbf{Task-state-aware message-passing}\quad Now, we have our task-aware message-passing layer and state-aware message-passing layer, we combine them as our task-state-aware message-passing layer, which can be represented as follows:
\begin{equation}
\small
\label{eq:task_state_message}
    \begin{aligned}
        \bm{h}^{l}_{i,\bm{x}_t} &= {\bm{W}^l_4} \cdot OneHot({\bm{x}_{t,i}}), \quad 
        \bm{h}^{l}_{j,\bm{x}_t} = {\bm{W}^l_5} \cdot OneHot({\bm{x}_{t,i}})\\
        \bm{\hat{h}}^l_{i,t} &= {\bm{W}^l_2} \cdot ({\bm{h}^{l}_{i,\bm{x}_t}} \odot {\bm{h}^l_{i,t}}), \quad 
        \bm{\hat{h}}^l_{j,t} = {\bm{W}^l_3} \cdot ({\bm{h}^{l}_{j,\bm{x}_t}} \odot {\bm{h}^l_{j,t}})\\
        \bm{g}^l_{i,t} &= \sigma (\bm{W}^l_6 \cdot (({\bm{h}^{l}_{i,\bm{x}_t}} \odot {\bm{h}^l_{i,t}}) \odot({\bm{W}_d} \cdot ReLU({\bm{h}_{C_t}}))))\\
        \bm{g}^l_{j,t} &= \sigma (\bm{W}^l_7 \cdot (({\bm{h}^{l}_{j,\bm{x}_t}} \odot {\bm{h}^l_{j,t}}) \odot({\bm{W}_s} \cdot ReLU({\bm{h}_{C_t}}))))\\
        \bm{h}^{l+1}_{i,t} &=ReLU({\bm{W}^l_1}{\bm{h}^l_{i,t}}+\sum_{j\in \mathcal{N}(i)} (\bm{\hat{h}}^l_{i,t} \odot \bm{g}^l_{i,t} - {\bm{\hat{h}}^l_{j,t}} \odot \bm{g}^l_{j,t}))
    \end{aligned}
\end{equation}
where $\bm{W}_1$, $\bm{W}_2$, $\bm{W}_3$, $\bm{W}_4$, $\bm{W}_5$, $\bm{W}_6$, $\bm{W}_7$ are learnable parameters at layer $l$, and $\bm{W}_d$ and $\bm{W}_s$ are learnable parameters shared across all layers. 
The idea of Equation \ref{eq:task_state_message} is to first scale the source nodes' representation and the target nodes' representation according to their states, then apply task-aware message-passing. 

\noindent \textbf{Task-state-aware GNN}\quad To encode the local graph structure for large $k$, we need to stack our task-state-aware message-passing layers with a large depth. 
However, GNN with a large depth often suffer from over-smoothing \cite{li2018deeper} issue. 
Given a $N$-layer task-state-aware GNN, we employ skip connection \cite{he2016deep} to connect the input of the initial GNN layer and the output of the last GNN layer to alleviate the effect of over-smoothing. 
Moreover, different subgraph sizes $k$ necessitate different sizes of local graph structures and, consequently, a different number of GNN layers. 
While this could be handled by a task-aware component within a task-state-aware message-passing framework, relying solely on the final GNN layer may introduce bias, particularly when the model has not been well-trained.
To better filter out redundant information, we extract the local graph structure from the last GNN layer according to the task representation using the gating mechanism. 
The final task-state-aware node representation is computed using a gated skip connection:
\begin{equation}
\small
\label{eq:task_state_gnn}
\begin{aligned}
    \bm{h}^0_{i,t} &= ReLU(\bm{W}_0 \cdot OneHot(L^T(v^T_i))) \\
    \bm{h}^N_{i,t} &= \text{GNN}(\bm{h}^0_{i,t}) \\
    \bm{g}_{i,t} &= \sigma(\bm{h}^N_{i,t} \odot({\bm{W}_c} \cdot ReLU({\bm{h}_{C_t}})))\\
    \bm{h}_{i,t} &= \bm{h}^0_{i,t} + \bm{h}^N_{i,t} \odot  \bm{g}_{i,t}
\end{aligned}
\end{equation}
where $h^0_{i,t}$ is the initial node representation extracted from the node label $L^T(v^T_i)$ using learnable parameter $\bm{W}_0$,
$h^N_{i,t}$ is the output of the last GNN layer, $\bm{W}_c$ is a learnable parameter, and $h_{i,t}$ is the final representation of node $v^T_i \in V^T$ at time t. 
Equation \ref{eq:task_representation}, \ref{eq:task_state_message} and \ref{eq:task_state_gnn} compose our task-state-aware GNN as illustrated in Figure \ref{fig:gnn}(a).

\subsubsection{Prediction Head} After encoding the local graph structure into node representation $\bm{h}_{i,t}$, we compute the Q-value / probability of each action according to its local graph structure's probability of capturing the most frequent subgraph in the global graph structure.

\noindent \textbf{Actor Prediction Head}\quad The prediction head of the Actor network computes the probability of each action for a given state as presented in Figure \ref{fig:gnn}(b).
The node representations $H_t$ are passed through a Multi-Head Attention Layer \cite{vaswani2017attention}, which compares the subgraphs captured by each node’s local graph structure with those captured by other nodes.
This process can be interpreted as estimating the ``frequency" of each subgraph within the global graph structure.

Next, we use multi-layer perceptron (MLP) to predict each node's probability of being selected:
\begin{equation}
\small
    \begin{aligned}
        H_t &= [\bm{h}_{0,t},\bm{h}_{0,t},...,\bm{h}_{|V^T|,t}]^T\\
        \hat{H}_t = \text{MHA}(H_t,H_t,H_t)&, \quad \hat{\bm{h}}_{a_t,t} = \hat{H}_{t}[a_t] \\
        \text{Mask}(\text{MLP}(\hat{H}_t))_{a_t} &= 
        \begin{cases}
            \text{MLP}(\hat{\bm{h}}_{a_t,t}) & \text{if }  a_t \in \mathcal{A}_t \\
            -1 \times 10^8 & \text{otherwise}
        \end{cases} \\
         \pi_\psi(\cdot | G^T,S_t,C_t) &= \text{Softmax}(\text{Mask}(\text{MLP}(\hat{H}_t))) \\
    \end{aligned}
\end{equation}
where $H_t$ denotes the node representation matrix, MHA denotes the Multi-head Attention Layer, $\hat{\bm{h}}_{a_t,t}$ is the representation of the node $v^T_{a_t}$ selected by action $a_t$, and $\pi_\psi(\cdot | G^T,S_t,C_t)$ denotes the Action network's output action probability distribution parameterized by $\psi$. 
Moreover, $\mathcal{A}_t$ represents the valid action space at time $t$, the Mask function indicates that Softmax is only computed on valid actions, invalid actions result in $0$ probability of being selected \cite{huang2020closer}.

\noindent \textbf{Critic Prediction Head}\quad Given the node representation matrix $H_t$, the Q-value prediction head of the Critic network computes the Q-value of taking each action with the Dueling network architecture \cite{wang2016dueling} as shown in Figure \ref{fig:gnn}(c):
\begin{equation}
\small
    \begin{aligned}
        \hat{Q}_\theta((G^T,S_t,C_t),a_t) &= V_{\theta}(H_t)+(A_{\theta}(H_t,a_t) - \frac{1}{|\mathcal{A}_t|}\sum_{a'_t \in \mathcal{A}_t}A_{\theta}(H_t,a'_t))
    \end{aligned}
\end{equation}
where $\hat{Q}_\theta((G^T,S_t,C_t),a_t)$ denotes the Q-value computed by the Critic prediction head parameterized by $\theta$, $V_{\theta}$ is the value network that computes the value of $(G^T,S_t,C_t)$, and $A_{\theta}$ is the action advantage network that computes the value of the action $a_t$. 

The action advantage network follows the same architecture as the Actor network:
\begin{equation}
\small
    \begin{aligned}
        \hat{H}_t = \text{MHA}(H_t,H_t,H_t)&,\quad\hat{\bm{h}}_{a_t,t} = \hat{H}_{t}[a_t]\\
        A_{\theta}(H_t,a_t) &= \text{MLP}(\hat{\bm{h}}_{a_t,t})
    \end{aligned}
\end{equation}
where $\hat{\bm{h}}_{a_t,t}$ denotes the representation of node $v^T_{a_t}$ selected by action $a_t$.

The value network first performs graph pooling to obtain the representation of $(G^T,S_t,C_t)$, then computes the value as follows:
\begin{equation}
\small
    \begin{aligned}
        \bm{h}_{(G^T,S_t,C_t)} &= \text{MLP}(\text{PMA}(H_t)) \\
        V_{\theta}(H_t) &= \text{MLP}(\bm{h}_{(G^T,S_t,C_t)})
    \end{aligned}
\end{equation}
where PMA is the Multi-head Attention Pooling layer \cite{lee2019set}, and $\bm{h}_{(G^T,S_t,C_t)}$ is the representation of $(G^T,S_t,C_t)$.

To assist the representation learning of the Critic network, we add a reward prediction head to the Critic network as proposed in DeepMDP \cite{gelada2019deepmdp}. 
The reward prediction head is represented as follows:
\begin{equation}
\small
    \begin{aligned}
        \hat{r}_\theta((G^T,S_t,C_t),a_t) = \text{MLP}(\text{CONCAT}[\bm{h}_{(G^T,S_t,C_t)},\hat{\bm{h}}_{a_t,t}])
    \end{aligned}
\end{equation}
where CONCAT denotes the concatenation operation. 

Due to the space limit, the detailed training algorithm of the Critic network and the Actor network is described in Appendix \ref{sec:training_detail}.

\subsection{Time Complexity}
\label{sec:complexity}
During inference, we select the action based on the output action distribution computed by the Actor network. 
The time complexity of task-state-aware GNN is $O(|V^T|\cdot d^2+|E^T|\cdot d)$, and the time complexity of the prediction head is $O(|V^T|^2\cdot d + |V^T| \cdot d^2)$. 
The overall time complexity of the computing action probabilities using the Actor network is $O(|V^T|\cdot d^2 + |V^T|^2\cdot d)$, since $|E^T| \leq |V^T|^2$. 
For a given subgraph size $k$, RLMiner takes $k$ steps to find the subgraph, and the overall time complexity of finding a $k$-sized subgraph is $O(k(|V^T|\cdot d^2 + |V^T|^2\cdot d))$, which is linear to $k$ without pruning the search space. 
Note that, the computation of subgraph frequency is not required during inference, therefore RLMiner has very low time complexity compared to traditional exact enumeration. 

\section{Experiments}
\label{sec:experiment}
In this section, we present the experimental results that demonstrate the superior performance and efficiency achieved by our RLMiner.

\begin{table}[t]
\caption{Dataset description}
\vspace{-2mm}
    \label{table:dataset}
    \centering
    \begin{tabular}{|c|c|c|c|}
    \hline
    Dataset& Number of Graphs & Avg $|V^T|$ &Avg $|E^T|$\\
    \hline
    ENZYMES&552&32.42&62.94\\
    \hline
    COX2&467&41.22&43.45 \\
    \hline
    BZR&405&35.75&38.36 \\
    \hline
    DHFR&756&42.43&44.54 \\
    \hline
    PROTEINS&910&41.62&77.39 \\
    \hline
    \end{tabular}
\end{table}

\subsection{Dataset}
We conduct experiments over real-world datasets obtained from TUDatasets
to evaluate RLMiner. 
For each dataset, we aim to find induced subgraphs of size $k \in \{5,6,7,8,9\}$, since the search space for subgraph size $k=3$ or $4$ is relatively small, and the exact enumeration failed to scale beyond $k > 9$. 
For simplicity, only connected graphs with $500 > |V^T| > 10$ are used as target graphs, and $80\%$ of target graphs are used for training, the remaining target graphs are used for evaluation. 
The statistics of the target graph datasets used in our experiments are summarized in Table \ref{table:dataset}.

\subsection{Baseline methods}
In order to demonstrate the performance and efficiency of RLMiner, the following methods are compared:

    \noindent\textbf{Exact Enumeration}: Since there is no known exact algorithm to directly find the most frequent induced subgraph of size $k$ in a single labeled graph with frequency measurement without DCP, we use Gaston\footnote{\url{https://liacs.leidenuniv.nl/~nijssensgr/gaston/index.html}} \cite{gaston} to enumerate all possible induced subgraphs, then use VF3\footnote{\url{https://github.com/MiviaLab/vf3lib}} \cite{carletti2017challenging} to count the frequency of each induced subgraph to find the most frequent one. 

    Gaston was designed for finding all the frequent subgraphs across a graph dataset with frequency higher than a given frequency threshold, where a subgraph is counted as $0$ or $1$ in a single target graph. Therefore, to enumerate all the induced subgraphs in a single target graph, we set frequency threshold to $1$.

    \noindent\textbf{T-FSM-serial}: T-FSM\footnote{\url{https://github.com/lyuheng/T-FSM}} \cite{yuan2023t} is an exact enumeration method for finding frequent subgraphs in a single labeled graph with $MNI$ higher than a given threshold. We use T-FSM-serial to find the induced subgraphs with the highest $MNI$, then use VF3 to count the frequency of each founded induced subgraphs to find the most frequent one, this is considered as an approximate method. 
    
    However, the official implementation of T-FSM-serial does not return the $MNI$ of each founded subgraph, and it is difficult to directly determine a proper threshold. Thus, we initially set the threshold to the highest possible $MNI$ and run T-FSM-serial, then iteratively reduce the threshold and re-run T-FSM-serial until we find the induced subgraphs with the highest $MNI$ for all subgraph sizes $k \in K$.

    \noindent\textbf{Rand}: We replace the Actor network of RLMiner by a random action sampler to sample $100$ induced subgraphs for a given target graph, then use VF3 to count the frequency of each sampled subgraphs to find the most frequent one, this serves as an approximate method. We repeat and average the results over $5$ seeds.

For RLMiner, the reward for training is also computed by VF3. Due to the space limit, the implementation detail of RLMiner is described in Appendix \ref{sec:implementation}. 

For all methods, we conduct experiments using Nvidia Geforce RTX3090 24GB and Intel i9-12900K with 128GB RAM.

\subsection{Evaluation Metrics}
We use the average approximation ratio over test target graphs to evaluate the performance of each method for each subgraph size $k$, the approximation ratio on a single test target graph $G^T$ is defined as $\frac{\textit{frequency }(G^S_k,G^T)}{\textit{frequency }({G^S_k}^*,G^T)}$, where $G^S_k$ is the $k$-sized induced subgraph found by the method under evaluation, and ${G^S_k}^*$ is the ground-truth induced subgraph found by the Exact Enumeration.

\begin{table*}[ht]

   \centering
    \begin{tabular}{|c|c|c|c|c|c|c|c|c|c|c|c|c|c|c|c|c|c|c|c|c|c|c|c|c|}
    \hline
         \multirow{2}{*}{\backslashbox{Method}{Dataset}}&\multicolumn{6}{c|}{ENZYMES}&\multicolumn{6}{c|}{COX2} \\
    \cline{2-13}
    &k=5&k=6&k=7&k=8&k=9&Total time (s)&k=5&k=6&k=7&k=8&k=9&Total time (s) \\
    \hline
    Exact Enumeration&1&1&1&1&1&31076.60&1&1&1&1&1&168.36\\
    \hline
    T-FSM-serial&0.701&0.604&0.554&0.542&0.533&122.25&0.850&0.486&0.722&0.486&0.633&\textbf{6.17}\\
    \hline
    Rand&\textbf{0.982}&\textbf{0.918}&0.767&0.630&0.500&373.25&0.998&0.976&0.910&0.775&0.711&287.55\\
    \hline
    \textbf{RLMiner (ours)}&0.913&0.884&\textbf{0.835}&\textbf{0.810}&\textbf{0.779}&\textbf{11.11}&\textbf{0.999}&\textbf{0.998}&\textbf{0.986}&\textbf{0.988}&\textbf{0.971}&9.67\\
    \hline
    \end{tabular}\\[10pt]
    
    \begin{tabular}{|c|c|c|c|c|c|c|c|c|c|c|c|c|c|c|c|c|c|c|c|c|c|c|c|c|}
    \hline
         \multirow{2}{*}{\backslashbox{Method}{Dataset}}&\multicolumn{6}{c|}{BZR}&\multicolumn{6}{c|}{DHFR} \\
    \cline{2-13}
   &k=5&k=6&k=7&k=8&k=9&Total time (s)&k=5&k=6&k=7&k=8&k=9&Total time (s) \\
    \hline
    Exact Enumeration&1&1&1&1&1&194.03&1&1&1&1&1&311.78\\
    \hline
    T-FSM-serial&0.866&0.618&0.776&0.596&0.693&\textbf{4.15}&0.747&0.403&0.552&0.426&0.530&17.85\\
    \hline
    Rand&0.996&0.973&0.901&0.745&0.649&243.77&0.994&0.955&0.915&0.820&0.749&472.05\\
    \hline
    \textbf{RLMiner (ours)}&\textbf{0.997}&\textbf{0.997}&\textbf{0.988}&\textbf{0.967}&\textbf{0.955}&8.23&\textbf{0.997}&\textbf{0.993}&\textbf{0.977}&\textbf{0.968}&\textbf{0.934}&\textbf{15.44}\\
    \hline
    \end{tabular}\\[10pt]

    \begin{tabular}{|c|c|c|c|c|c|c|c|c|c|c|c|c|c|c|c|c|c|c|c|c|c|c|c|c|}
    \hline
         \multirow{2}{*}{\backslashbox{Method}{Dataset}}&\multicolumn{6}{c|}{PROTEINS} \\
    \cline{2-7}
   &k=5&k=6&k=7&k=8&k=9&Total time (s)\\
    \hline
    Exact Enumeration&1&1&1&1&1&39024.57\\
    \hline
    T-FSM-serial&0.769&0.689&0.686&0.621&0.625&358.60\\
    \hline
    Rand&\textbf{0.9865}&\textbf{0.919}&0.808&0.656&0.564&2897.67\\
    \hline
    \textbf{RLMiner (ours)}&0.907&0.876&\textbf{0.839}&\textbf{0.799}&\textbf{0.769}&\textbf{18.10}\\
    \hline
    \end{tabular}

    \caption{Average approximation ratio on test target graphs. Total time indicates the total running time of finding the most frequent induced subgraph from size $k=5$ to $k=9$.}
    \vspace{-6mm}
    \label{table:overall_result}
\end{table*}

\subsection{Result}

    \noindent \textbf{\textit{RQ1.}} \textit{How does RLMiner perform compare to other methods in terms of performance and inference time?}

    Table \ref{table:overall_result} shows each method's average approximation ratio on test target graphs for each subgraph size $k$. Since Exact Enumeration and T-FSM-serial enumerate subgraphs from $k=1$ up to $k=9$ together in a single run, we report the total running time of finding the most frequent induced subgraph from $k=5$ up to $k=9$. 
    
    For smaller subgraph size $k=5$ and $6$, the search space is relatively small, Rand can achieve high approximation ratio by simple sampling. However, the search space increases exponentially as $k$ increases, RLMiner has a much better performance than other baseline methods for larger $k$ across all datasets, even for the harder target graphs (ENZYMES, PROTEINS). For simpler target graphs (COX2, BZR, DHFR), RLMiner can almost perfectly find the most frequent induced subgraph for all $k$.
    
    Moreover, RLMiner enjoys a much shorter running time compared to exact enumeration as shown in Table \ref{table:overall_result}. 
    Especially for harder target graphs in ENZYMES and PROTEINS dataset, the running time of Exact Enumeration is more than $2000\times$ longer compared to the running time of RLMiner. 
    This is because counting the frequency of the subgraph is NP-hard, and the Exact Enumeration requires counting a significant number of subgraphs that grows exponentially with $k$. 
    Our proposed RLMiner does not require the enumeration of subgraphs and avoids the counting of subgraph frequency during inference, thus RLMiner has a much shorter running time.

    \vspace{0.5em}\noindent \textbf{\textit{RQ2.}} \textit{Can we find the most frequent induced subgraph by finding the induced subgraph with the highest \textit{MNI}?}

    From Table \ref{table:overall_result}, it is obvious that using the induced subgraph with the highest \textit{MNI} to approximate the ground-truth most frequent induced subgraph reduces the running time significantly. However, the approximation ratio becomes very low, even for small $k$. Moreover, the approximation ratio tends to be unstable, e.g., the approximation ratio of subgraph size $k=6$ is much lower than the approximation ratio of $k=7$ for COX2 dataset. 
    Therefore, pruning the search space using frequency measure with DCP, such as the most popular measurement \textit{MNI}, cannot approximate the ground-truth most frequent induced subgraph in a proper manner.

    \begin{table}[h]
        \centering
        \begin{tabular}{|c|c|c|c|c|c|c|}
            \hline
             \multirow{2}{*}{\backslashbox{Size}{Dataset}}&\multicolumn{3}{c|}{ENZYMES}&\multicolumn{3}{c|}{PROTEINS}\\
             \cline{2-7}
             &Eq 2&Eq 3&Eq 4&Eq 2&Eq 3&Eq 4\\
             \hline
             k=5&0.878&0.913&0.890&0.825&0.907&0.842\\
             \hline
             k=6&0.807&0.884&0.842&0.748&0.876&0.801\\
             \hline
             k=7&0.760&0.835&0.771&0.680&0.839&0.739\\
             \hline
             k=8&0.734&0.810&0.745&0.633&0.799&0.658\\
             \hline
             k=9&0.685&0.779&0.691&0.585&0.769&0.614\\
             \hline
        \end{tabular}
        \caption{Approximation ratio of RLMiner on unseen test target graphs for different reward settings.}
        \vspace{-5mm}
        \label{table:reward_eval}
    \end{table}
    \begin{table}[h]
        \centering
    \begin{tabular}{|c|c|c|}
        \hline
        \backslashbox{Size}{Dataset}&ENZYMES&PROTEINS\\
        \hline
        k=5&0.989&0.976\\
        \hline
        k=6&0.984&0.963\\
        \hline
        k=7&0.977&0.936\\
        \hline
        k=8&0.964&0.912\\
        \hline
        k=9&0.950&0.865\\
        \hline
    \end{tabular}
        \caption{Approximation ratio of RLminer on training target graphs.}
        \vspace{-7mm}
        \label{table:reward_train}
    \end{table}
    \vspace{0.5em}\noindent \textbf{\textit{RQ3.}} \textit{How does reward settings affect the performance of RLMiner?}
    
    Table \ref{table:reward_eval} shows the maximum average approximation ratio achieved by RLMiner with different reward settings on test target graphs during training for harder datasets ENZYMES and PROTEINS (the full curve is visualized in Appendix \ref{Appx:curve} Fig \ref{fig:reward_eval}). Normalizing the reward by the frequency of the ground-truth subgraph (Eq \ref{eq:reward2}) results in superior and more stable performance, compared to using a reward without normalization (Eq \ref{eq:reward1}) or a reward normalized by the target graph size, edge density, and subgraph size (Eq \ref{eq:reward3}). Equation \ref{eq:reward2} normalizes the reward into $(0,1]$ based on each specific target graph and subgraph size $k$, rather than using a monotonic function. This also aligns RLMiner's objective (maximizing the cumulative reward) with the evaluation metric, thus Eq \ref{eq:reward2} leads to superior performance on challenging unseen test target graphs. 
    
    Normalizing the reward using a monotonic function or not normalizing it at all (Eq \ref{eq:reward1}) can introduce bias when the frequency of the ground-truth induced subgraph for unseen test target graphs significantly differs from the frequency of induced subgraphs for training target graphs. This approach also causes the objective of RLMiner to shift towards maximizing the total frequency of the founded induced subgraphs across the target graph dataset. Consequently, RLMiner tends to focus more on target graphs with potentially higher subgraph frequencies, which does not align with our evaluation metric. This misalignment leads to poor and fluctuating performance on test target graphs.
    
   In practice, obtaining the ground-truth most frequent induced subgraph for each training target graph and subgraph size $k$ is often infeasible. Eq \ref{eq:reward3} simply approximates the ground-truth frequency using the target graph's property and subgraph size $k$, it results in better performance than not normalizing the reward. However, there remains a significant performance gap compared to normalizing by the actual ground-truth frequency. Approximating the ground-truth frequency by solely using the target graph size, edge density, and multiplying by subgraph size $k$ is insufficient, as the number of possible subgraphs grows exponentially with $k$.

   Moreover, Table \ref{table:reward_train} shows the maximum average approximation ratio achieved by RLMiner on training target graphs (the full curve is visualized in Appendix \ref{Appx:curve} Fig \ref{fig:reward_train}). For harder target graphs, RLMiner with reward Eq \ref{eq:reward2} can find frequent subgraphs on training target graphs with very high approximation ratio for all subgraph size $k$. Also, the training time of RLMiner is shorter than the running time of exact enumeration for harder target graphs, and the total number of subgraphs sampled ($\text{number of epochs} \times \text{number of training graphs} \times |K|$) from the training target graphs during training is very small compared to the exponential search space of the problem (total number of subgraphs in the training target graphs). If the ground-truth frequency can be accurately approximated, RLMiner could serve as a much more efficient alternative to exact enumeration when high accuracy is required and the search time is not severely constrained. Therefore, approximating the frequency of ground-truth most frequent $k$-sized induced subgraph can be a valuable direction for future research.

   Due to the space limit, more detailed ablation study is provided in Appendix \ref{sec:ablation}.

\section{Conclusion}
\label{sec:conclusion}
In this paper, we present a novel multi-task reinforcement learning-based framework RLMiner to find the most frequent induced subgraph with size $k$. 
By formulating the problem as a Markov Decision Process and leveraging a multi-task reinforcement learning approach integrated with our proposed task-state-aware Graph Neural Network, RLMiner effectively addresses the computational challenges posed by traditional exact enumeration methods, which can approximately find the most frequent induced subgraph with time complexity linear to $k$. 
Our extensive experiments on real-world datasets demonstrate that RLMiner not only accurately identifies high-frequency subgraphs but also significantly reduces the running time.


\bibliographystyle{ACM-Reference-Format}
\bibliography{acmart-primary/acmart}

\appendix
\section{Training Details}
\label{sec:training_detail}
\begin{algorithm}[t]
\small
    \caption{RLMiner Training Procedure}
    \renewcommand{\algorithmicrequire}{\textbf{Input:}}
    \begin{algorithmic}[1]
        \label{algo}
        \REQUIRE Training graphs $[G^T_0,G^T_1,...,G^T_n]$, subgraph sizes $K$, target network soft update factor $\tau$;
        \STATE Initialize network parameters $\theta_1, \theta_2, \psi$, temperature $\alpha$;
        \STATE Initialize target network parameters $\bar{\theta}_1 \leftarrow \theta_1, \bar{\theta}_2 \leftarrow \theta_2$;
        \FOR{each $k \in K$}
            \STATE Initialize replay buffer $D_k \leftarrow \emptyset$;
        \ENDFOR
        \FOR{each epoch}
            \FOR{each training graph $G^T_i$}
                \FOR{each $k \in K$}
                    \WHILE{not terminate}
                        \STATE $a_t \sim \pi_\psi(a_t | G^T_i,S_t,C_t)$;
                        \STATE $\bm{x}_{t+1,{a_t}} \leftarrow 1$;
                        \STATE $C_{t+1} \leftarrow (k-t-1,k)$;
                        \STATE $r_t \leftarrow r((G^T_i,S_t,C_t),a_t)$;
                        \STATE $d_t \leftarrow 1$ if terminates, else $d_t \leftarrow 0$;
                        \STATE $D_k \leftarrow D_k \cup \{((G^T_i,S_t,C_t),a_t,r_t, (G^T_i,S_{t+1},C_{t+1})),d_t\}$;
                    \ENDWHILE
                    \FOR{each gradient step}
                        \STATE Initialize batch $B \leftarrow \emptyset$;
                        \FOR{each $k \in K$}
                            \STATE $B \leftarrow B \cup$\{sampled batch $B_k \sim D_k$\};    
                        \ENDFOR
                        \STATE Update $\theta_1, \theta_2, \psi, \alpha$ over Equation \ref{loss:theta}, \ref{loss:psi}, \ref{loss:alpha} using $B$;
                            \STATE $\bar{\theta}_1 \leftarrow \tau \cdot \theta_1 + (1-\tau) \cdot \bar{\theta}_1, \bar{\theta}_2 \leftarrow \tau \cdot \theta_2 + (1-\tau) \cdot \bar{\theta}_2$;
                    \ENDFOR
                \ENDFOR
            \ENDFOR
            \STATE $\bar{\theta}_1 \leftarrow \theta_1, \bar{\theta}_2 \leftarrow \theta_2$;
        \ENDFOR
    \end{algorithmic}
\end{algorithm}

We follow the objectives as described in the discrete version of SAC \cite{christodoulou2019soft} to update the Actor network and the Critic network, and with double Q-learning trick. The overall training algorithm of RLMiner is presented in Algorithm \ref{algo}.

\noindent The objective of the $i$-th Critic network's Q-value prediction head is to minimize:
\begin{equation}
\small
    \begin{aligned}
        J_{Q}(\theta_i) = E_{((G^T,S_t,C_t),a_t)\sim D}[&\frac{1}{2}(\hat{Q}_{\theta_i}((G^T,S_t,C_t),a_t)\\ 
        &- Q((G^T,S_t,C_t),a_t))^2]
    \end{aligned}
\end{equation}
with 
\begin{equation}
\small
    \begin{aligned}
        Q((G^T,S_t,C_t),a_t) = r((G^T,S_t,C_t),a_t) + \gamma E_{a_{t+1} \sim \pi_\psi}[\\
        \underset{i=1,2}{\min}(\hat{Q}_{\bar{\theta}_i}((G^T,S_{t+1},C_{t+1}),a_{t+1}))\\
        -\alpha \log \pi_\psi(a_{t+1} | G^T,S_{t+1},C_{t+1})]
    \end{aligned}
\end{equation}
where $D$ is a replay buffer, $\hat{Q}_{\bar{\theta}_i}$ is the $i$-th target critic network updated periodically according to $\hat{Q}_{{\theta}_i}$, $\gamma$ is the discount factor, $\alpha$ is the learnable temperature parameter, and $(G^T,S_{t+1},C_{t+1})$ is incurred by taking $a_t$ from $(G^T,S_t,C_t)$.

\noindent The objective for the reward prediction head of the $i$-th Critic network is to minimize:
\begin{equation}
\small
\begin{aligned}
    J_{reward}(\theta_i) =  E_{((G^T,S_t,C_t),a_t)\sim D}[(&\hat{r}_{\theta_i}((G^T,S_t,C_t),a_t) - \\ &r((G^T,S_t,C_t),a_t))^2]
\end{aligned}
\end{equation}
where $r((G^T,S_t,C_t),a_t)$ is the true reward. 

\noindent The overall objective of the $i$-th Critic network is to minimize:
\begin{equation}
\small
\label{loss:theta}
    J(\theta_i) = J_Q(\theta_i) + J_{reward}(\theta_i)
\end{equation}

\noindent The objective of the Actor network is to minimize:
\begin{equation}
\small
\label{loss:psi}
    \begin{aligned}
        J(\psi) = E_{((G^T,S_t,C_t),a_t)\sim D}[E_{a_t \sim \pi_\psi}[\alpha \log (\pi_\psi (a_t | G^T,S_{t},C_{t})) \\
        - \underset{i=1,2}{\min}(\hat{Q}_{{\theta}_i}((G^T,S_{t},C_{t}),a_{t}))]]
    \end{aligned}
\end{equation}

\noindent The objective of the temperature parameter $\alpha$ is to minimize:
\begin{equation}
\small
\label{loss:alpha}
    \begin{aligned}
        J(\alpha) &= E_{a_t \sim \pi_\psi}[\alpha (-\pi_\psi(a_t | G^T,S_{t},C_{t})-\mathcal{\bar{H}}_t)] \\
        \mathcal{\bar{H}}_t &= 0.6 \cdot -\log (\frac{1}{|\mathcal{A}_t|})
    \end{aligned}
\end{equation}
where $\mathcal{\bar{H}}_t$ is the target entropy computed based on the number of valid actions. 

\section{Implementation Details}
\label{sec:implementation}
For both the Actor network and the Critic network of RLMiner, the hidden dimension is set to $256$, we adopt $9$ layers of task-state-aware message passing layers, and the number of heads is set to $4$ for MHA and PMA layers. 

For the training of RLMiner, we employ Prioritized Experience Replay (PER) \cite{schaul2015prioritized} as our replay buffers, where each subgraph size $k$ has an individual replay buffer with size $10^{-6}$, the prioritization exponent of PER is set to $0.2$ and $\beta$ of PER is set to 0.6. Moreover, we set discount factor $\gamma$ to $0.99$, initial $\alpha$ to $1$ and target network soft update factor $\tau$ to $10^{-2}$. We adopt Adam optimizer with learning rate $2.5 \times 10^{-4}$. Before training, we randomly sample states from training target graphs and store into replay buffers for $50$ epochs. 
Next, for COX2, BZR and DHFR datasets, we train RLMiner with $75$ epochs. For ENZYMES dataset, we train RLMiner with $150$ epochs since ENZYMES dataset is considered to be harder target graphs. For PROTEINs dataset, although it is also considered to be harder target graphs, we train RLMIner with $75$ epochs to limit the training time.
During training, we perform $2$ gradient steps to update the parameters of RLMiner after each termination of episode, where we sample $8$ experiences from each replay buffer for each gradient step. In addition, although we only aim to find subgraphs with size $k=5$ to $k=9$, we still train RLMiner with subgraph size from $k=3$ up to $k=9$, this allows RLMiner to learn basic knowledge from simpler tasks instead of directly learning from harder tasks.

\begin{figure*}[ht]
    \centering
    \includegraphics[scale=0.35]{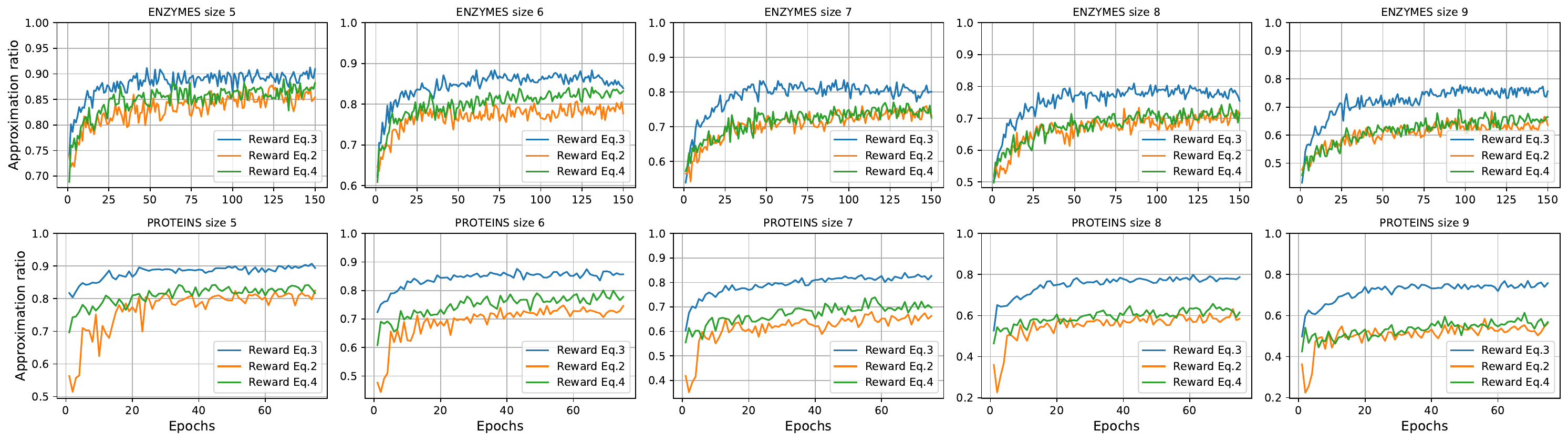}
    \caption{Approximation ratio of RLMiner on unseen test target graphs during training for different reward settings.}
    \label{fig:reward_eval}
    \end{figure*}

    \begin{figure*}[ht]
    \centering
    \includegraphics[scale=0.35]{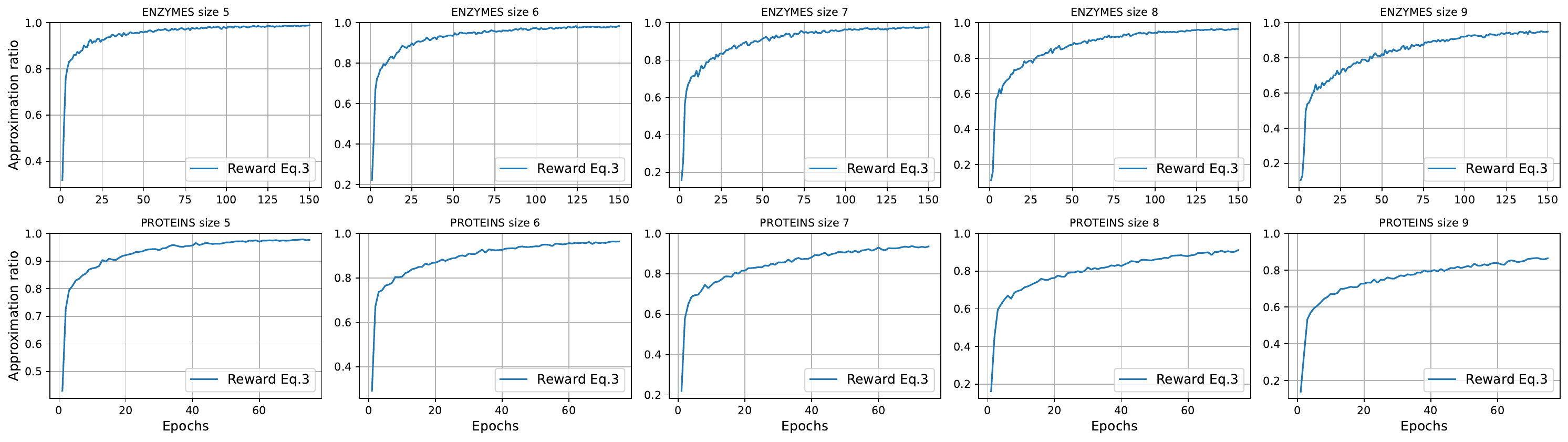}
    \caption{Approximation ratio of RLminer on training target graphs during training.}
    \label{fig:reward_train}
    \end{figure*}
    \begin{figure*}[ht]
    \centering
    \includegraphics[scale=0.35]{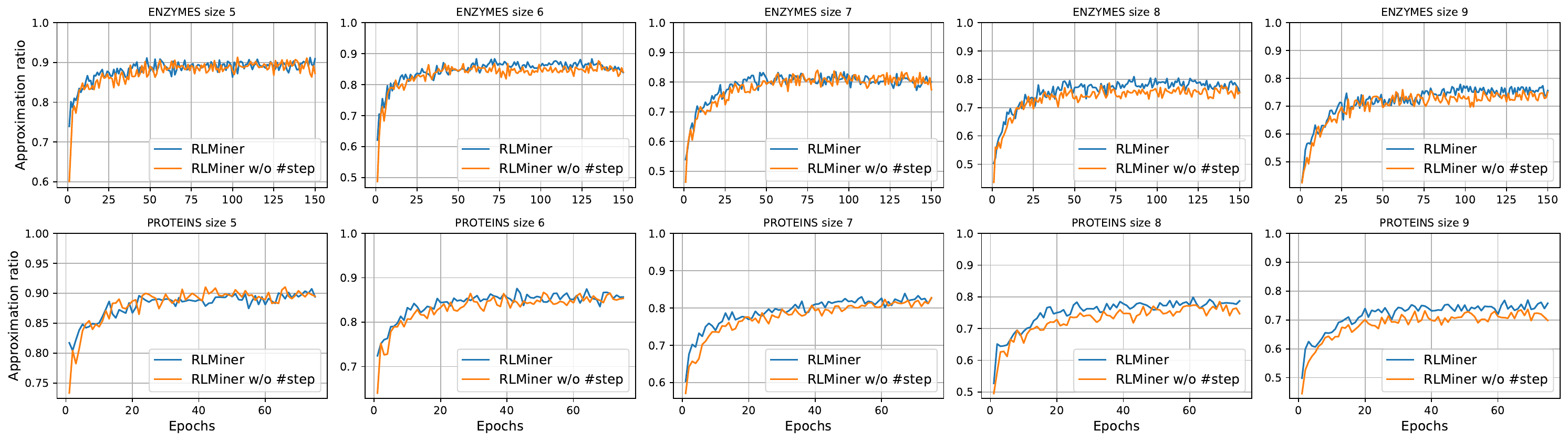}
    \caption{Approximation ratio of RLMiner on unseen test target graphs during training for different task information settings.}
    \label{fig:taskinfo}
    \vspace{-3mm}
    \end{figure*}
    \begin{figure*}[ht]
    \centering
    \includegraphics[scale=0.35]{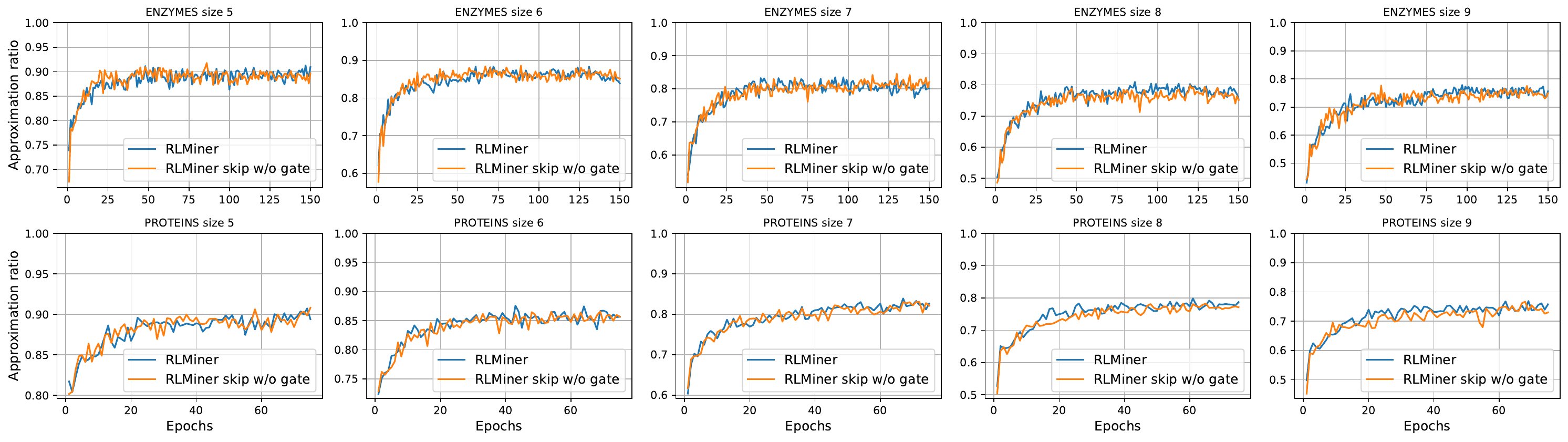}
    \caption{Approximation ratio of RLMiner on unseen test target graphs during training for different skip connection.}
    \label{fig:skipnogate}
    \vspace{-3mm}
    \end{figure*}

    \begin{figure*}[ht]
    \centering
    \includegraphics[scale=0.35]{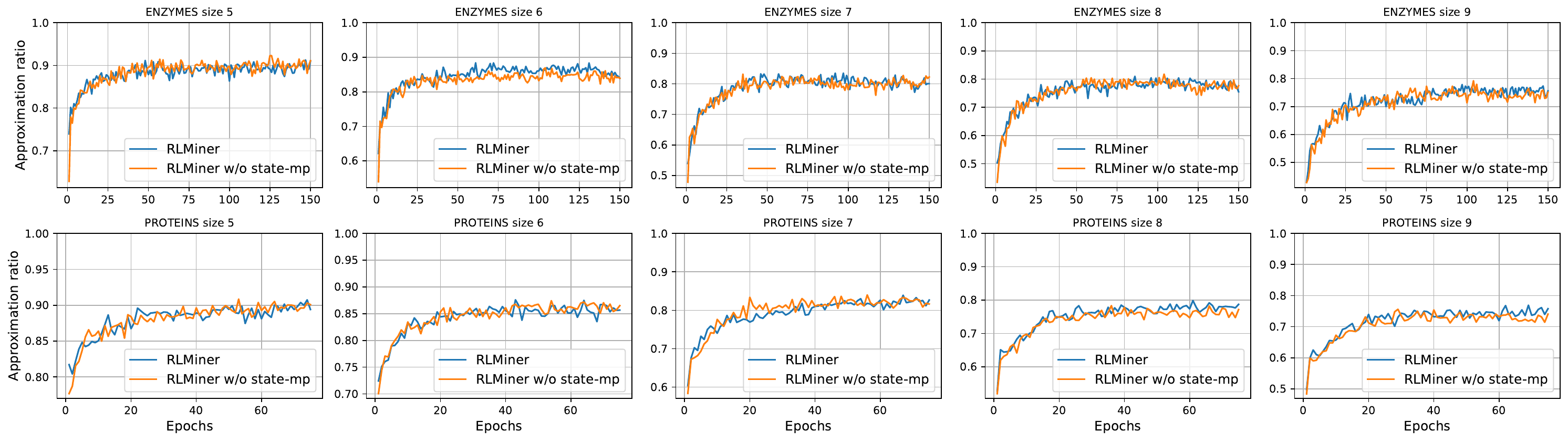}
    \caption{Approximation ratio of RLMiner on unseen test target graphs during training for different state settings.}
    \label{fig:stateaware}
    \vspace{-3mm}
    \end{figure*}
    \begin{figure*}[ht]
    \centering
    \includegraphics[scale=0.35]{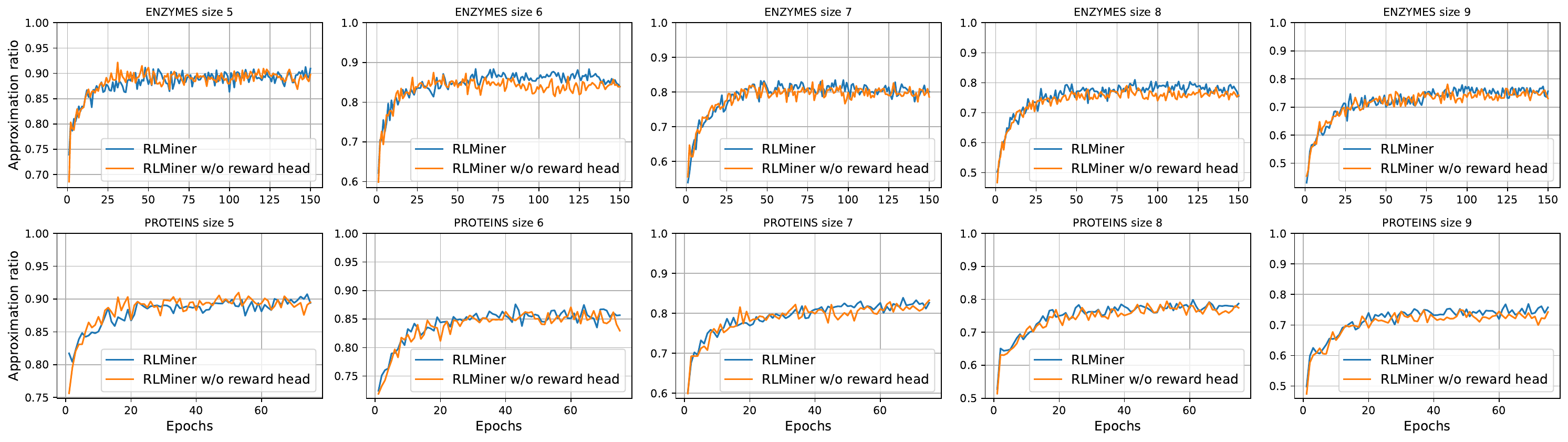}
    \caption{Approximation ratio of RLMiner on unseen test target graphs during training with and without reward prediction head.}
    \label{fig:rewardhead}
    \vspace{-3mm}
    \end{figure*}
\section{Additional Experimental Results}
\subsection{Training Curve Visualization}
\label{Appx:curve}
Figure \ref{fig:reward_eval} shows the curve of average approximation ratio achieved by RLMiner with different reward settings on unseen test target graphs during training. Figure \ref{fig:reward_train} shows the curve of average approximation ratio achieved by RLMiner on training target graphs during training.

\subsection{Ablation Study}
    \label{sec:ablation}
   How does each key component of the Actor network and the Critic network influence the performance of RLMiner?

    Since ENZYMES and PROTEINS dataset were demonstrated to be more challenging, we perform ablation study on these two datasets.

     \noindent \textbf{Task information}\quad In Section \ref{sec:mdp}, we include the counting of remaining steps as part of the task information $C_t$ for RLMiner. To measure the importance of task information, we create a variant model (RLMiner w/o \#step) that excludes the remaining steps from $C_t$.
     From Figure \ref{fig:taskinfo}, we can see that including the remaining steps for RLMiner generalized better on unseen test graphs in early and middle stages, and performed better in late stage for larger $k$. For successive states $S_t$ and $S_{t+1}$, the only difference between $\bm{x}_t$ and $\bm{x}_{t+1}$ is $\bm{x}_{t,{a_t}}$, and the state difference is only reflected through the state-aware component of task-state-aware GNN, thus making RLMiner difficult to distinguish the successive states, especially when RLMiner is not well-trained in early stage. Including the counting of remaining steps as part of the task information allows RLMiner to recognize the state difference not only by the state-aware component, but also by the task-aware component.
     Moreover, as $k$ increases, finding the most frequent induced subgraph becomes more challenging. The counting of remaining steps provides additional task-related information to assist RLMiner for harder tasks.

    \noindent \textbf{Task-aware gated skip connection}\quad
    To measure the importance of gated skip connection as described in Equation \ref{eq:task_state_gnn}, we create a variant model (RLMiner skip w/o gate) that employs skip connection without gating mechanism, the average approximation ratio on test graphs are shown in Figure \ref{fig:skipnogate}. As mentioned in Section \ref{sec:approach}, different subgraph size $k$ requires different size of local graph structures, directly using the final GNN layer may introduce bias if the model is not well-trained. Although it is possible to be filtered out by the task-aware message-passing, using the gated skip connection could filter out redundant information in a more direct way. Therefore, RLMiner with gated skip connection can achieve better performance for tasks that are harder to be well-trained (subgraph size of $8$ and $9$).
    \noindent \textbf{State-aware message-passing}\quad 
    To evaluate the effectiveness of the proposed state-aware message-passing, we create a variant model (RLMiner w/o state-mp), which concatenates $\bm{x}_t$ with the node label matrix of target graph and pass into task-aware GNN. As shown in Figure \ref{fig:stateaware}, both approaches exhibit similar performance on test target graphs for smaller values of $k$ , while RLMiner with state-aware message-passing has better performance for larger values of $k$. As explained in Section \ref{sec:representation}, concatenating $\bm{x}_t$ with the node label matrix may introduce bias when comparing the extracted local graph structures to determine "frequency." However, nodes in the available action space are always unselected, meaning they share the same state value in $\bm{x}_t$. Consequently, this concatenation does not significantly impact the ranking of each action's probability or value. Also, for smaller values of $k$, only a few nodes are selected, which does not lead to much bias when comparing local graph structures, and the bias increases as $k$ becomes larger. Therefore, there is no significant performance gap between two approaches for smaller values of $k$, but RLMiner with state-aware message passing performed better for larger value of $k$.

    \noindent \textbf{Reward prediction head}\quad
    To measure the significance of adding a reward prediction head to the Critic network, we create a variant model (RLMiner w/o reward head) that removes the reward prediction head from the Critic network. As shown in Figure \ref{fig:rewardhead}, initially, removing the reward prediction head results in comparable or even superior performance on test target graphs compared to RLMiner with the reward prediction head. However, from middle stage of training, the performance of RLMiner without reward prediction head starts to become worse and unstable for the majority of $k$ comparing to RLMiner with reward prediction head. This is because adding an additional prediction head makes the network more complex and difficult to optimize in early stage. As training progresses, the reward prediction head introduces informative regulation signal on the representation learning of the Critic network, thereby leading to a more stable and better performance.

\end{document}